\documentclass{emulateapj}
\usepackage{colordvi}
\usepackage{epsfig}
\usepackage{graphicx}

\usepackage{amssymb}
\usepackage{threeparttable}
\usepackage{tabularx}

\newcommand{\Ha}{\hbox{{\rm H}\kern 0.1em$\alpha$}}
\newcommand{\Hb}{\hbox{{\rm H}\kern 0.1em$\beta$}}
\newcommand{\MgII}{\hbox{{\rm Mg}\kern 0.1em{\sc ii}}}
\newcommand{\CIV}{\hbox{{\rm C}\kern 0.1em{\sc iv}}}
\newcommand{\NeV}{\hbox{[{\rm Ne}\kern 0.1em{\sc v}]}}
\newcommand{\OII}{\hbox{[{\rm O}\kern 0.1em{\sc ii}]}}
\newcommand{\NeIII}{\hbox{[{\rm Ne}\kern 0.1em{\sc iii}]}}
\newcommand{\OIII}{\hbox{[{\rm O}\kern 0.1em{\sc iii}]}}
\newcommand{\NII}{\hbox{[{\rm N}\kern 0.1em{\sc ii}]}}
\newcommand{\SII}{\hbox{[{\rm S}\kern 0.1em{\sc ii}]}}

\newcommand{\lsfr}{log(SFR/M$_{\odot}$~y$^{-1}$)}
\newcommand{\lmass}{log(M$_{\star}$/M$_{\odot}$)}
\newcommand{\lmdyn}{log(M$_{\rm dyn}$/M$_{\odot}$)}
\newcommand{\mdyn}{M$_{\rm{dyn}}$}
\newcommand{\mbar}{M$_{\rm{bar}}$}
\newcommand{\mgas}{M$_{\rm{gas}}$}
\newcommand{\mstar}{M$_{\star}$}

\usepackage{color}
\definecolor{citeRGB}{rgb}{0,0.1,0.7}
\usepackage[hyperfootnotes=true,naturalnames=true,letterpaper,pdfstartview=FitH,pdfpagemode=UseNone,colorlinks=true,citecolor=citeRGB]{hyperref}

\shorttitle{ALMA CO kinematics of a compact SFG}
\shortauthors{Barro et al.}
\slugcomment{To be submitted to ApJ letters}

\begin{document}

\title{Spatially resolved kinematics in the central 1~kpc of a compact
  star-forming galaxy at $\lowercase{z}\sim2.3$ from ALMA CO
  observations}


\author{G. Barro\altaffilmark{1,2}
M. Kriek\altaffilmark{2},
P. G.~P\'{e}rez-Gonz\'{a}lez\altaffilmark{3}
T. Diaz-Santos\altaffilmark{4},
S. H. Price\altaffilmark{5},
W. Rujopakarn\altaffilmark{6,7,8},
V. Pandya\altaffilmark{9}, 
D. C.~Koo\altaffilmark{9}, 
S. M.~Faber\altaffilmark{9},
A. Dekel\altaffilmark{10},
J. R.~Primack\altaffilmark{11},
D. D.~Kocevski\altaffilmark{12}}

\altaffiltext{1}{University of the Pacific, USA}
\altaffiltext{2}{University of California Berkeley, USA}
\altaffiltext{3}{Universidad Complutense de Madrid, Spain}
\altaffiltext{4}{Universidad Diego Portales, Chile}
\altaffiltext{5}{Max Planck Institute for Extraterrestrial Physics, Germany}
\altaffiltext{6}{Chulalongkorn University, Thailand}
\altaffiltext{7}{National Astronomical Research Institute of Thailand}
\altaffiltext{8}{Kavli Institute for Physics and Mathematics, Japan}
\altaffiltext{9}{University of California Santa Cruz, USA}
\altaffiltext{10}{The Hebrew University, Israel}
\altaffiltext{11}{Santa Cruz Institute for Particle Physics, USA}
\altaffiltext{12}{Colby College, USA}


\slugcomment{To be submitted to the Astrophysical Journal Letters} 
\slugcomment{Last edited: \today}
\label{firstpage}
\begin{abstract}

We present high spatial resolution (FWHM$\sim$0\farcs14) observations
of the CO($8-7$) line in GDS-14876, a compact star-forming galaxy at
$z=2.3$ with total stellar mass of \lmass$=10.9$. The spatially
resolved velocity map of the inner $r\lesssim1$~kpc reveals a
continous velocity gradient consistent with the kinematics of a
rotating disk with $v_{\rm rot}(r=1\rm kpc)=163\pm5$~km~s$^{-1}$ and
$v_{\rm rot}/\sigma\sim2.5$. The gas-to-stellar ratios estimated from
CO($8-7$) and the dust continuum emission span a broad range, $f^{\rm
  CO}_{\rm gas}=M_{\rm gas}/M_{\star}=13-45\%$ and $f^{\rm cont}_{\rm
  gas}=50-67\%$, but are nonetheless consistent given the
uncertainties in the conversion factors. The dynamical modeling yields
a dynamical mass of \lmdyn$=10.58^{+0.5}_{-0.2}$ which is lower, but
still consistent with the baryonic mass, $\log$(M$_{\rm bar}$=\mstar +
M$^{\rm CO}_{\rm gas}$/M$_{\odot}$)$=11.0$, if the smallest CO-based
gas fraction is assumed. Despite a low, overall gas fraction, the
small physical extent of the dense, star-forming gas probed by
CO($8-7$), $\sim3\times$ smaller than the stellar size, implies a
strong relative concentration that increases the gas fraction up to
$f^{\rm CO, 1\rm kpc}_{\rm gas}\sim 85\%$ in the central 1~kpc. Such a
gas-rich center, coupled with a high star-formation rate,
SFR$\sim$~500~M$_{\odot}$~yr$^{-1}$, suggests that GDS-14876 is
quickly assembling a dense stellar component ({\it bulge}) in a strong
nuclear starburst. Assuming its gas reservoir is depleted without
replenishment, GDS-14876 will quickly ($t_{\rm depl}\sim27$~Myr)
become a compact quiescent galaxy that could retain some fraction of
the observed rotational support.



\end{abstract}

\section{Introduction}

Compact star-forming galaxies (SFGs) are frequently referred to as a
population of massive, strongly star-forming galaxies at $z\gtrsim2$
whose small sizes and high stellar mass concentrations (e.g.,
\citealt{wuyts11b}; \citealt{barro13}) closely resemble those of
typical quiescent galaxies of the same mass and redshift (e.g.,
\citealt{daddi05}; \citealt{trujillo07}). These galaxies have been
identified in sizable numbers and their properties: small stellar
sizes, steep radial mass profiles and obscured SFR properties have all
been confirmed by multiple studies (e.g., \citealt{barro14a};
\citealt{dokkum15}). Moreover, NIR spectroscopic follow-up has allowed
a characterization of the kinematic and dynamical properties of their
ionized gas from the analysis of rest-frame optical emission lines
(e.g., \citealt{barro14b}; \citealt{dokkum15};
\citealt{nelson15}). These initial results revealed high integrated
velocity dispersions ($\sigma\gtrsim200$~km~s$^{-1}$) and large
dynamical masses, roughly consistent with their stellar masses, which
imply relatively low gas (and dark matter) fractions, and short
depletion times. All this evidence is consistent with the evolutionary
picture in which compact SFGs are in a short-lived starburst phase,
triggered by a dissipative event, which leads to the rapid formation
of a compact core and subsequent quenching into a compact quiescent
galaxy (e.g.; \citealt{zolotov15}; \citealt{wellons15}).

\begin{figure*}[t]
\centering
\includegraphics[width=9cm,angle=0.]{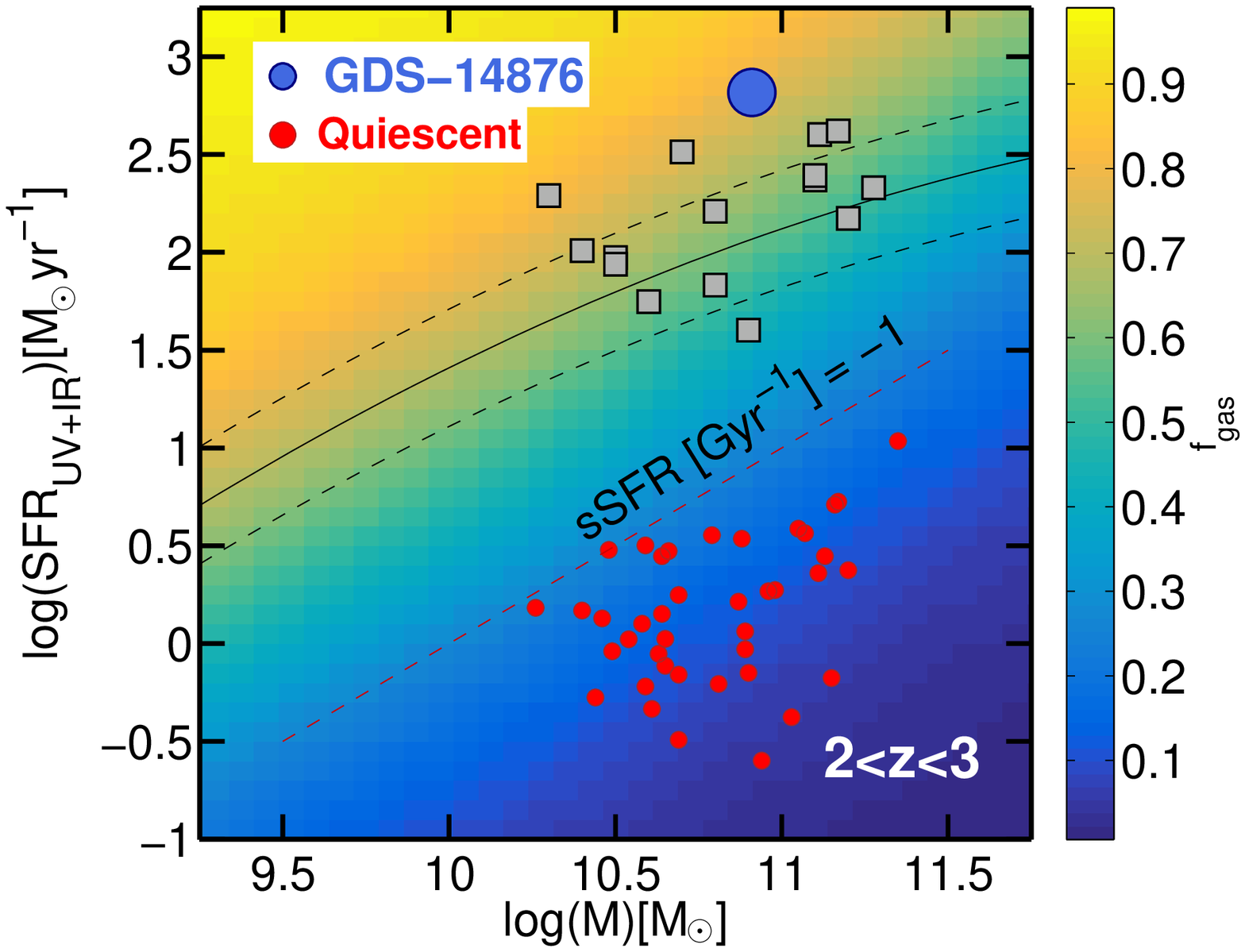}
\hspace{1cm}
\includegraphics[width=7.3cm,angle=0.]{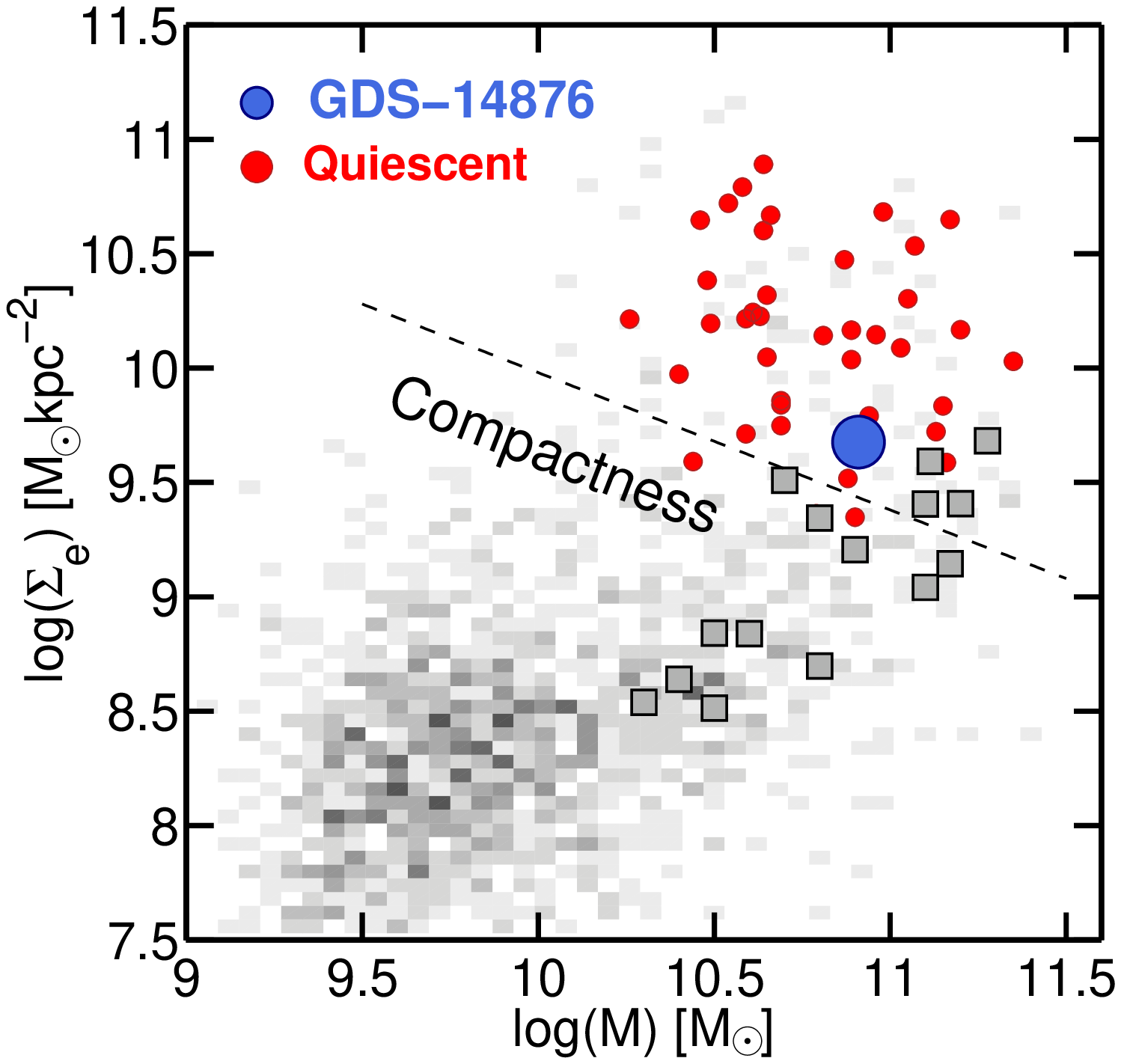}
\caption{\label{sfrms} {\it Left:} Logarithm of SFR vs. stellar mass
  for GDS-14876 (blue circle) and other massive SFGs at $z\sim2$ (grey
  squares) observed with ALMA and HST, drawn from \citet{spilker16},
  \citet{rujopakarn16} and \citet{popping17}. The black lines show the
  locus and $\times5$ limits (solid and dashed) of the SFR sequence at
  $z=2.25$ from \citet{whitaker14}. The background is colored by the
  predicted gas fraction determined from the empirical prediction of
  \citet{genzel15}. GDS-14876 lies above the SFR sequence and is
  expected to have a large $f_{\rm gas}\sim80\%$. The red line and
  circles show the location of the quiescent population at the same
  redshift, selected by low sSFR$<-1$~Gyr$^{-1}$. {\it Right:}
  Logarithm of the effective mass surface density vs. stellar mass for
  galaxies in CANDELS-GDS at $z\sim2$ (greyscale) and all galaxies
  from the left panel. The dashed line indicates the {\it compactness}
  selection criterion of \citet{barro17}.}
\end{figure*}

Nonetheless, tension with the dynamical constraints emerged when
further spectroscopic follow-up of compact SFGs revealed that at least
$20\%$ have dynamical masses that are up to $10\times$ lower than
their stellar masses \citep{dokkum15}. The most likely explanation for
such large discrepancies are the uncertainties on the dynamical
modeling assumptions. For example, the line-of-sight inclination, the
ratio between ordered and random motions of the gas (i.e., the amount
of rotational support, $v_{\rm rot}/\sigma$), the extent of gas
profile relative to the stellar mass distribution, or the aperture
corrections to scale the measurements within the slit to either
galaxy-wide or effective ($r=r_{\rm e}$) values can all contribute to
the observed difference (e.g., \citealt{dokkum15}; \citealt{price16}).

A way to reduce these uncertainties is obtaining emission line
velocity maps with similar or better spatial resolution than the
stellar mass maps derived from Hubble Space Telescope (HST) data
(e.g., \citealt{wuyts12}). These resolved maps can trace the kinematic
properties of the gas and allow a more precise dynamical modeling by
comparing the gas and stellar mass profiles at a similar scale. The
high spatial resolution of sub-mm spectroscopy with Atacama Large
Millimeter/sub-millimeter Array (ALMA) and the Karl G. Jansky Very
Large Array (VLA) are a perfect match for this analysis. Sub-mm
observations are insensitive to the dust obscuration that heavily
attenuates optical emission lines, and provide not only resolved
kinematics from CO and carbon lines (e.g., \citealt{tadaki17};
\citealt{popping17}), but also far IR continuum detections to
characterize the SFR and the baryonic content (i.e., gas and stars) of
the galaxies (e.g.; \citealt{scoville16}; \citealt{rujopakarn16}).

This work presents CO J=$8-7$ line and sub-mm continuum observations
of a compact SFG at z = 2.3 using ALMA. From the joint analysis of the
high spatial resolution HST/ACS, HST/WFC3 and ALMA continuum and CO
line imaging, we simultaneously characterize the spatial distribution
and kinematics of the molecular gas. Throughout this paper, we quote
magnitudes in the AB system, assume a \cite{chabrier} initial mass
function, and adopt the following cosmological parameters:
($\Omega_{M}$,$\Omega_{\Lambda}$,$h$) = (0.3, 0.7, 0.7).

\section{Target selection and Observations}

\begin{figure*}[t]
\centering
\includegraphics[width=7.2cm,angle=0.]{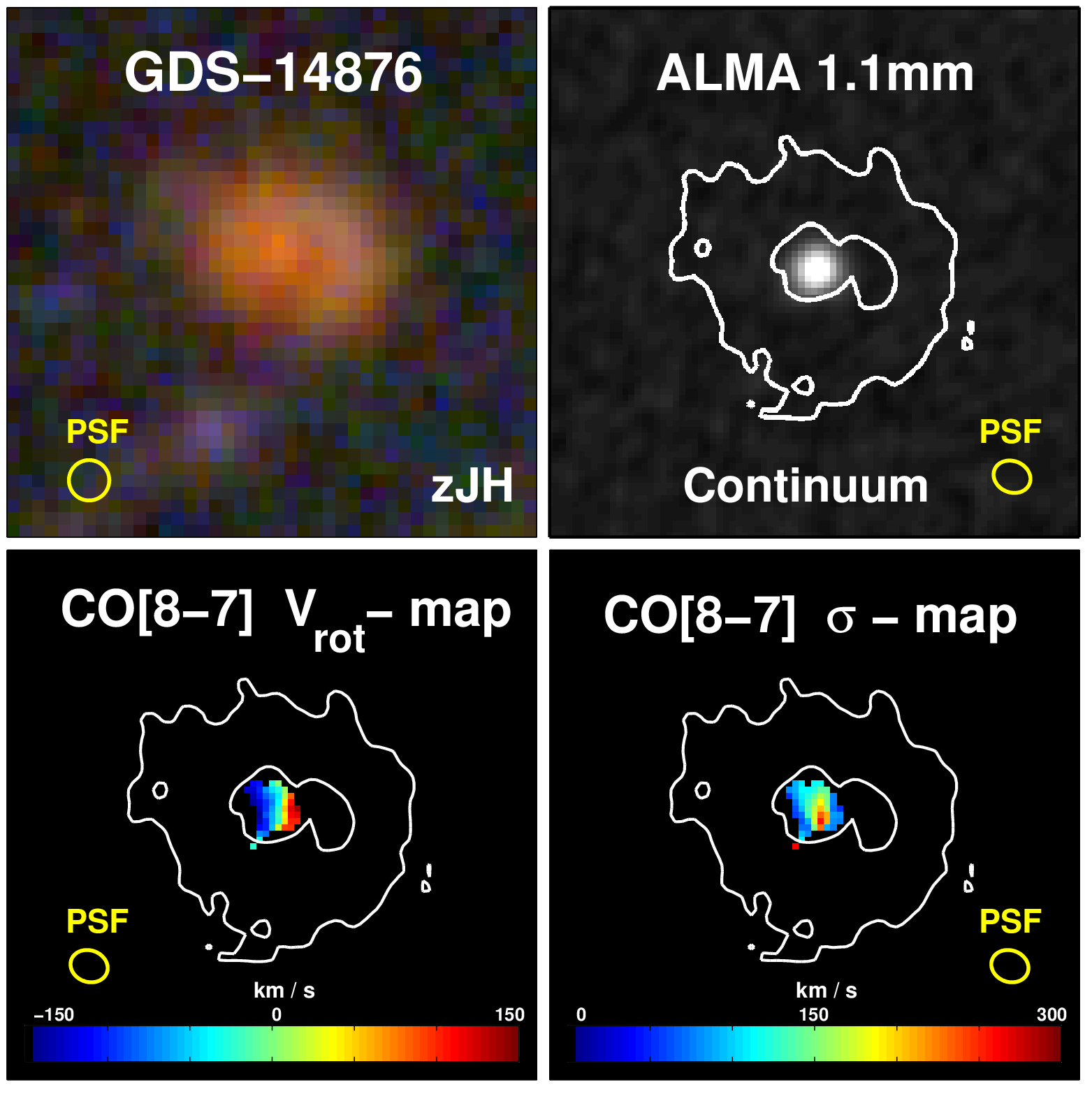}
\hspace{2cm}
\includegraphics[width=7.0cm,angle=0.]{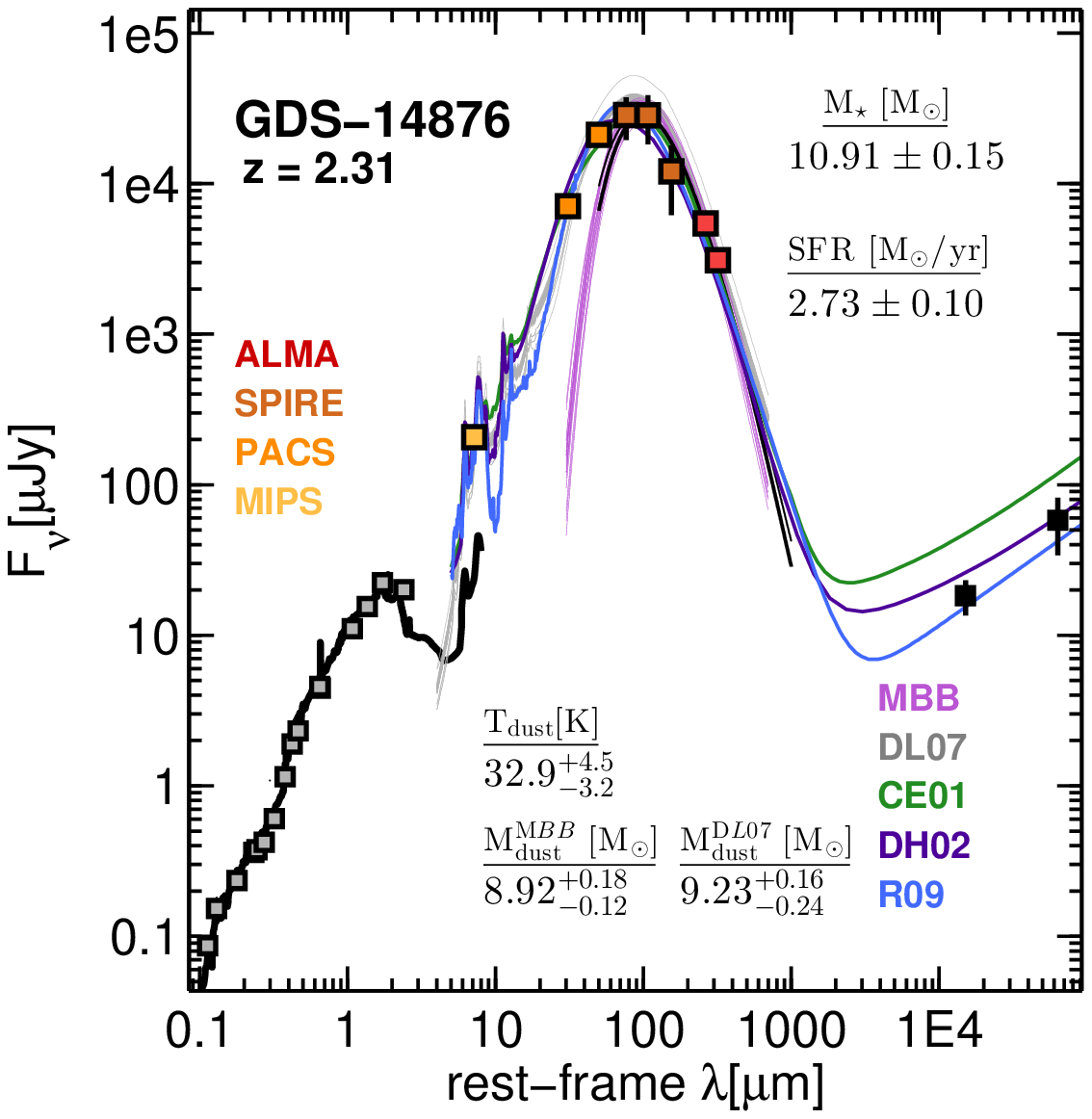}
\caption{\label{images} {\it Left:} From top to bottom,
  $2\farcs5\times2\farcs5$ (20$\times$20 kpc) images of GDS-14876 in
  WFC3/F160W and ALMA~870~$\mu$m, and CO($8-7$) rotation velocity and
  dispersion maps at the same physical scale (with the F160W contours
  shown in white). {\it Right:} UV-to-FIR SED of GDS-14876. The black
  line show the best-fit BC03 stellar population model for the
  photometry up to 8$\mu$m rest-frame (gray squares). The
  orange-to-red and black squares show the mid-to-far IR photometry
  and radio data. The green, purple and blue lines show the best-fit
  dust emission models from the libraries of \citet{ce01},
  \citet{dh02} and \citet{rieke09}. The grey and pink regions depict
  300 models drawn from the posterior probability distribution of the
  fit to the \citet{dl07} models and to a MBB model, respectively. The
  median values and confidence intervals for M$_{\rm dust}$ and
  $T_{\rm dust}$ are indicated.}
\end{figure*}

The galaxy analyzed in this paper is drawn from the compact SFGs
sample in the CANDELS (\citealt{candelsgro}) GOODS-S region presented
in \citet{barro14a, barro16b}. The panels in Figure~\ref{sfrms}
summarize the selection criteria in SFR and {\it
  compactness}. Figure~\ref{images} shows the UV to near-IR spectral
energy distribution (SED), which includes extensive multi-band data
ranging from U to 8$\mu$m \citep{guo13}.  Furthermore, we include
far-IR data from {\it Spitzer} MIPS \citep{pg08b}, {\it Herschel} PACS
and SPIRE from the GOODS-Herschel survey (\citealt{elbaz11}), and VLA
21~cm and 5~cm maps (\citealt{kellerman08};
\citealt{rujopakarn16}). The stellar population properties are
determined by fitting the optical \& NIR SED using FAST \citep{fast},
assuming \cite{bc03} stellar population synthesis models, an
exponentially-declining star formation history, and the
\cite{calzetti} dust law with attenuation $0<A_{V}<4$, yielding
\lmass$=$10.9. The SFR was determined in \citet{barro16b} from a
combination of rest-frame UV and IR SFR indicators and modeling the
far-IR emission with dust-emission templates, yielding \lsfr=2.7.

The sub-mm observations of GDS-14876 were taken as part of an ALMA
cycle-3 campaign (ID: 2015.1.00907.S; PI: G. Barro) to study CO
emission lines in compact SFGs at $z=2-3$. The observations were
conducted on 2016-09-17 in band 7 using four spectral windows in the
largest bandwidth mode. The on-source integration time was $51$~min in
an extended array configuration, C39-7 (shortest and longest baselines
were 15.1~m and 3.1~km, respectively). The water vapor during the
observations was PWV = 0.5 mm. Flux, phase, and band-pass calibrators
were also obtained, for a total time of $\sim2$~hr. All the
correlators were set to a bandwidth of 1.875 GHz covering 128
channels. The reference spectral window (spw0) was centered at
278.57~GHz to target the CO(8-7) emission line, assuming the redshift
derived from optical lines ($z = 2.309$). The other spectral windows
were centered at 276.91, 288.91, and 290.91~GHz. These spectral
windows were used to observe the band 7 continuum of the target.

We use the CASA software \citep{mcmullin07} to process and clean the
data. We use the {\tt tclean} task with natural weighting for the {\tt
  u-v} visibility plane. This resulted in a synthetic beam size with
an average angular resolution of FWHM~$=0\farcs14\times0\farcs12$
($1.15\times0.98$~kpc), with a major-axis position angle (P.A.) of
$70^{\circ}$. The r.m.s. noise of the observations is 0.19 mJy/beam
for the CO(8-7) line, measured in 20 km $s^{-1}$ channel bins, and
$\sigma=28$~$\mu$Jy/beam or $1.5$~mJy/arcsec$^{2}$ for the continuum,
measured on the clean continuum maps excluding the frequency range of
the CO line.

\section{Structural Properties}

Figure~\ref{images} shows the WFC3/ACS color-composite image of the
galaxy as well as the ALMA 1.1~mm continuum emission and CO(8-7)
velocity maps.  We measure the structural properties of GDS-14876 in
the F160W image, which at $z=2.3$ traces the rest-frame optical
emission, using \texttt{GALFIT} \citep{galfit} assuming a single
two-dimensional S\'ersic profile. We obtain size and S\'ersic index
values of $r_{e}=2.3\pm0.1$~kpc and $n=0.6$. The total (99\%)
isophotal size is $r_{\rm tot}=8.25\pm0.04$~kpc (outer white contours
in Figure~\ref{images}).

Following \citet{barro16b}, we use \texttt{GALFIT} and the synthetic
PSF of the ALMA beam to compute the size and S\'ersic index of the
continuum emission, obtaining values of $r_{e}=0.74\pm0.04$~kpc,
$n=1.6$. To measure the structural properties of the CO emission line
region, we subtract the continuum flux from the image using CASA's
\texttt{uvsubcont} task and then collapse the resulting data-cube in
velocity space using a $\pm250$~km/s bin around the central wavelength
of the CO[8-7] line. Lastly, we run \texttt{GALFIT} on the resulting
image, obtaining values of $r_{e}=0.67\pm0.05$~kpc and $n=0.9$.

Both the continuum and emission line sizes are $\sim3\times$ smaller
than the rest-frame optical size (Figure~\ref{images}). This is
consistent with previous results based on dust continuum measurements
(e.g., \citealt{barro16b}; \citealt{tadaki15};
\citealt{rujopakarn16}), which suggest that the compact CO and far-IR
emission trace a strong nuclear starburst. This compact burst
contrasts with the typical inside-growth of star-forming galaxies in
which gas profiles are more extended than the stellar distribution
(e.g., \citealt{nelson15}). We note however that given the high order
CO transition, it is possible that ALMA only detects the region where
$J=8-7$ can be excited, i.e., a dense and strongly ionized region in
the center.



\section{Dust and gas masses}

\subsection{Continuum-based measurement}

We fit the mid-to-FIR SED using different libraries of dust emission
templates (e.g., \citealt{ce01}, \citealt{dh02}) to estimate the total
IR luminosity and SFR (see Figure~\ref{images}). In addition, we fit
to the models by \citet[][DL07]{dl07} and to a set of modified black
body models (MBB, e.g., \citealt{casey12}) to estimate the dust
temperature and dust mass of the galaxy. The best-fit models and the
corresponding confidence intervals are computed by exploring the
parameter space using the Python Markov-Chain Monte Carlo package {\tt
  emcee} \citep{mcmc}.  The MBB fit assumes an average dust emissivity
of $\kappa=1$~cm$^{2}$g$^{-1}$ at 850~$\mu$m (e.g., \citealt{dunne03};
\citealt{scoville16}) and a range of $\beta=1.5-2.5$. Both estimates
of the dust mass are roughly consistent within the errors.  The DL07
models yield a value of $\log(M^{\rm DL07}_{\rm
  dust}/M_{\odot})=9.2\pm0.3$, while the MBB models with best-fit
values $\beta\sim1.5$ and T=33~K provide a slightly lower value of
$\log(M^{\rm MBB}_{\rm dust}/M_{\odot})=8.9\pm0.2$. Based on these
M$_{\rm dust}$ values, we estimate the molecular gas content using the
gas-to-dust ratio by assuming $\delta_{\rm GDR}$M$_{\rm dust}$$=M_{\rm
  H2} + M_{\rm HI}\sim M_{\rm H2}$. The value of $\delta_{\rm GDR}$
depends primarily on the metallicity of the galaxy (e.g.,
\citealt{sandstrom13}). Here we assume the typical value for solar
metallicity, $\delta_{\rm GDR}\sim100$.  This leads to gas masses of
$\log(M^{\rm DL07}_{\rm gas}/M_{\odot})=11.2\pm0.3$ and $\log(M^{\rm
  MBB}_{\rm gas}/M_{\odot})=10.9\pm0.2$~dex. Lower values of the
metallicity would yield larger gas masses.

\subsection{CO-based measurement}

We estimate the gas mass from the CO($8-7$) line luminosity using
\begin{equation}
L'_{\rm CO}[{\rm K~km~s^{-1}pc^{2}}]=3.25\times10^{7} (S_{\rm CO} \Delta v) \frac{D^{2}_{L}}{(1+z)^{3}\nu^{2}_{\rm obs}}
\end{equation}
where $S_{\rm CO}\Delta v$ is the line flux and $D_{L}$ is the
luminosity distance. We obtain $\log~L'_{\rm CO,
  8-7}=9.65\pm0.17$~dex.  From $L'_{\rm CO}$, we estimate the total
mass of molecular hydrogen as $\log(M_{\rm
  gas}/M_{\odot})=\log(\alpha_{\rm CO}L'_{\rm CO, J}/R_{\rm J,1}$),
where $R_{\rm J,1}=[S_{J}/S(1-0)]/J^{2}$ is the conversion factor to
correct for the lower Rayleigh-Jeans brightness temperature of the
$J=8-7$ transition relative to $1-0$, and $\alpha_{\rm CO}$ is the
CO-to-H2 conversion factor.  For the $J=8-7$ transition the conversion
factor is relatively unconstrained, e.g., $R_{8,1}=0.9$
\citep{bradford09}, $R_{8,1}=0.01-0.4$ (\citealt{danielson13};
\citealt{kamenetzky16}). Here we adopt a conservative value of
$R_{8,1}=0.30\pm0.20$, where the more sub-thermally excited is the gas
($R_{8,1}\lll1$), or the lower is the ISM density, the higher is the
gas mass (see e.g., \citealt{daddi15}). The value of $\alpha_{\rm CO}$
also depends on the physical conditions of the ISM. Adopting
$\alpha_{\rm CO} =0.8$ M$_{\odot}$ (K km s$^{-1}$ pc$^{2}$)$^{-1}$,
the typical value for nearby ultra-luminous infrared galaxies and SMGs
(e.g., \citealt{tacconi08}), we obtain $\log(M^{\rm SMG}_{\rm
  gas}/M_{\odot})=10.08^{+0.30}_{-0.18}$~dex. If we use the larger
Milky Way CO-to-H$_{2}$ conversion, $\alpha_{\rm CO} =4.3$ M$_{\odot}$
(K km s$^{-1}$ pc$^{2}$)$^{-1}$, the resulting gas mass is
$\log(M^{\rm MW}_{\rm gas}/M_{\odot})=10.81^{+0.30}_{-0.18}$~dex.

In summary, different indicators yield values of the dense,
star-forming gas mass that may differ up to $\sim1$~dex, nonetheless,
given the wide range of modeling assumptions and uncertainties in the
conversion factors, these values are still consistent. In Section \S 6
we discuss which of these estimates is more consistent with the
dynamical constraints. The estimates based on the CO line provide a
lower limit and imply gas fractions of $f^{\rm CO}_{\rm
  gas}\sim13^{+10}_{-4}$\% to $45^{+17}_{-10}$\% for the SMG and MW
gas-to-mass conversions, respectively. The dust continuum based values
computed with either the MBB or DL07 models yield higher gas fractions
of $f^{\rm cont}_{\rm gas}=50\pm11$\% to $67\pm15$\%, respectively.

Note that these fractions refer to the total, integrated masses. The
spatially resolved gas-to-stellar ratio within the region detected in
CO ($r\lesssim1$kpc) is much larger, even for the relatively modest
CO-based gas masses, $f^{\rm CO, 1kpc}_{\rm gas}=85\%$. This value
indicates that GDS-14876 has a gas-rich nuclear region that fuels the
similarly compact starburst detected in the dust continuum
emission. Assuming no replenishment, the depletion time of this burst
is very short: $t_{\rm depl}=SFR/M^{\rm CO, SMG}_{\rm
  gas}=27\pm12$~Myr.

\begin{figure}[t]
\centering
\includegraphics[width=4.25cm,bb=35 41 459 474]{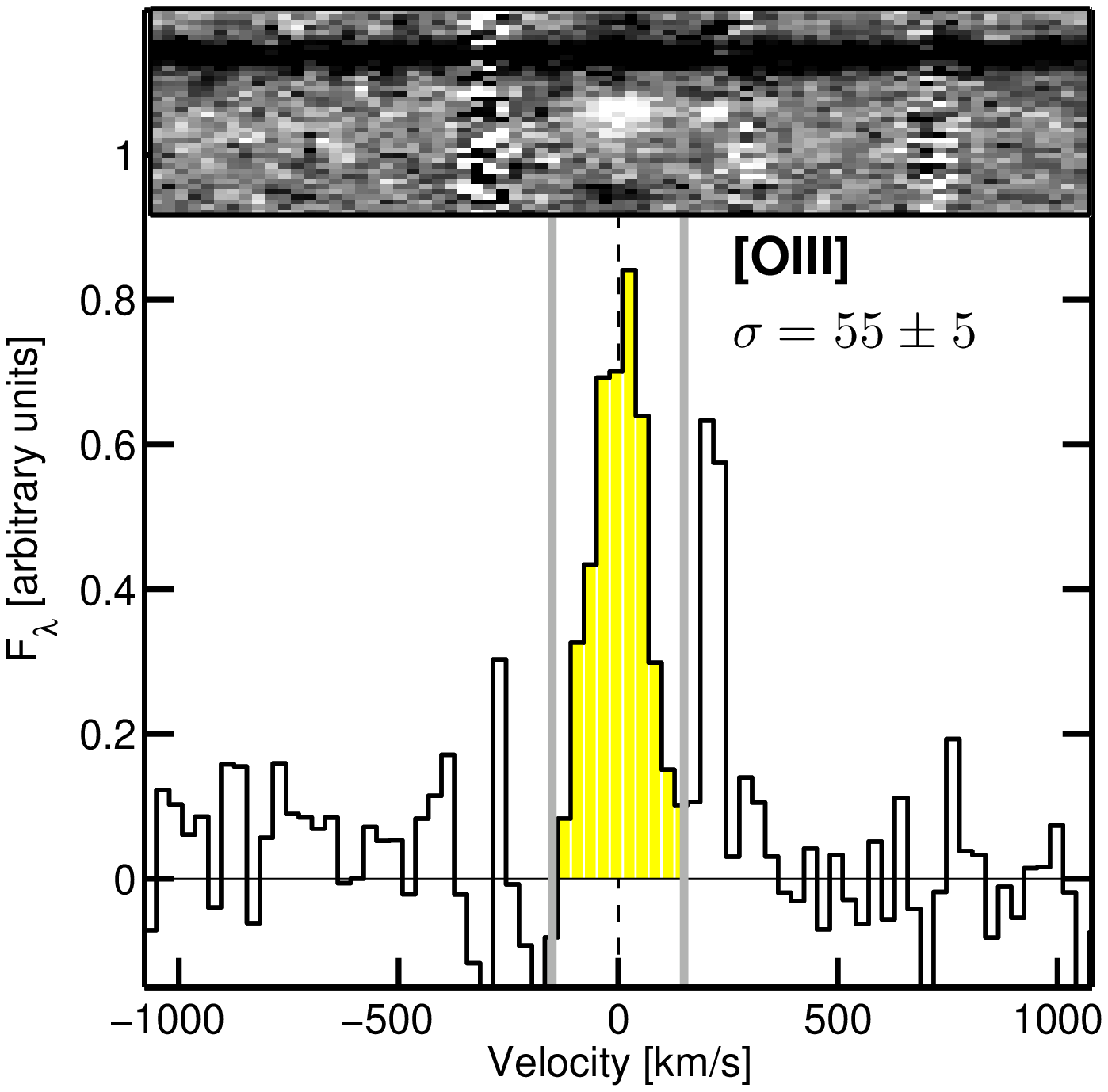}
\includegraphics[width=4.25cm,angle=0.]{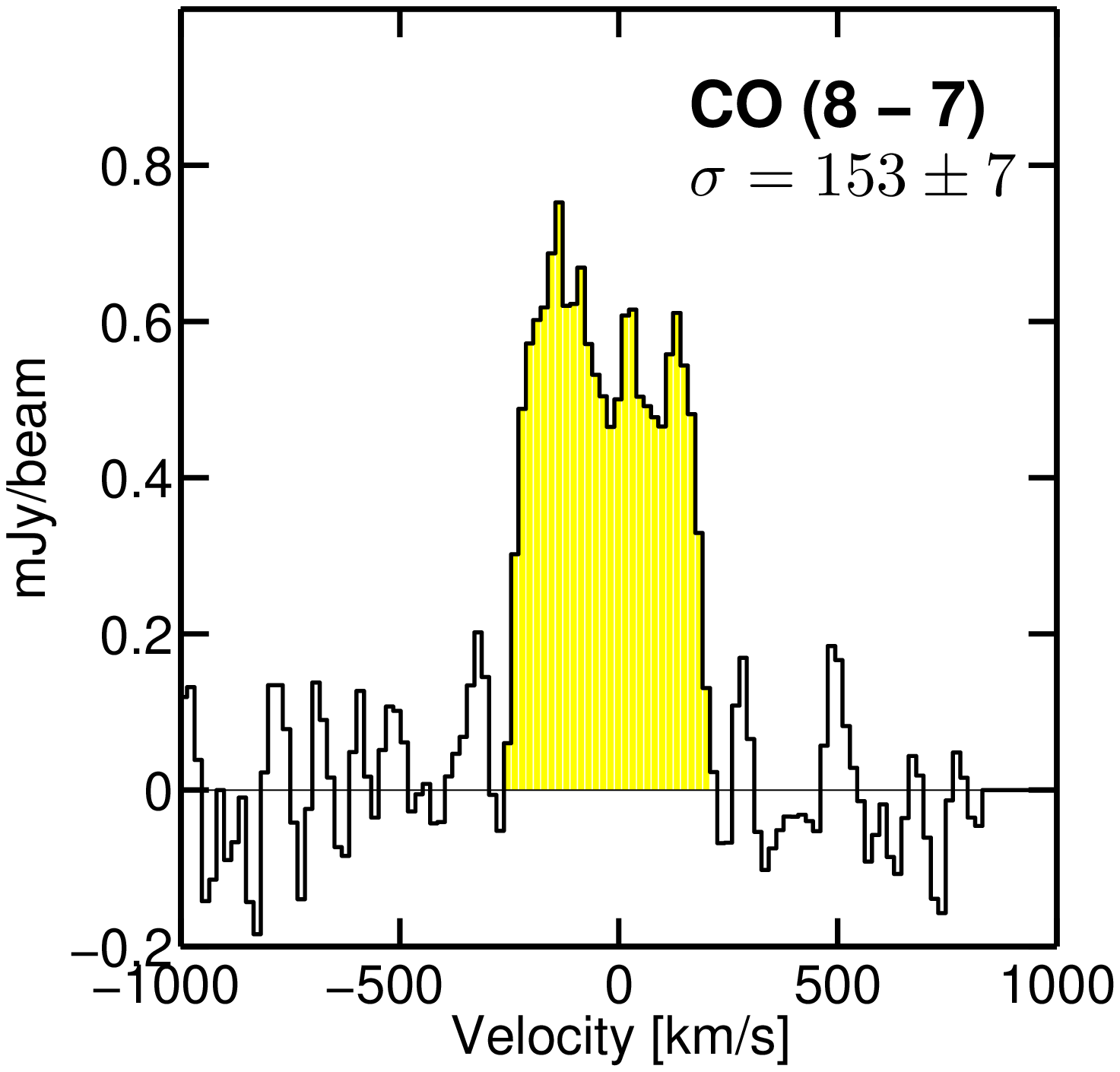}
\caption{\label{lines} {\it Left:} MOSFIRE H-band spectra of GDS-14876
  showing the 2D and collapsed 1D profile of the [OIII] emission
  line. The grey lines indicate $\pm$150~km~s$^{-1}$ for reference.
  {\it Right:} Continuum-subtracted spectral profile of the CO(8-7)
  line extracted from the ALMA image using a 0\farcs3 diameter
  circular aperture. At face-value, the dispersion of the CO line,
  measured from a single Gaussian fitting, is $\sim3\times$ larger
  than that of the optical [OIII] line.}
\end{figure}

\begin{figure*}[t]
\centering
\includegraphics[width=18.0cm]{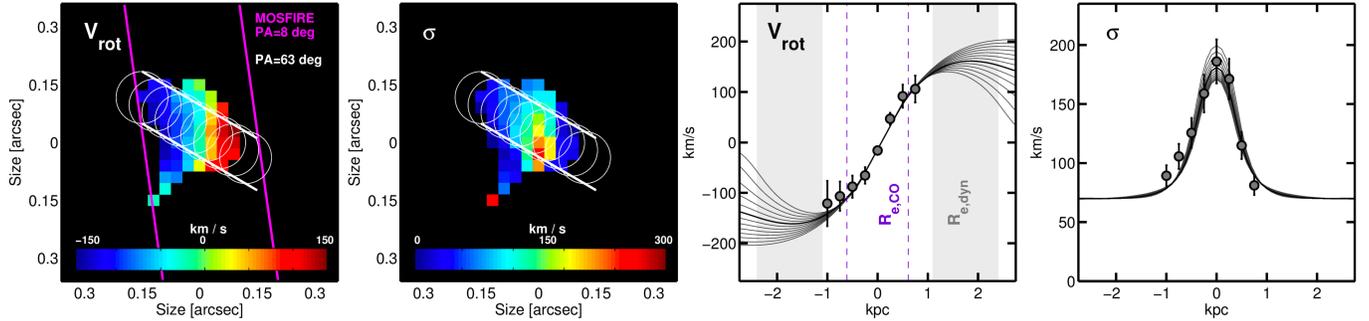}
\caption{\label{maps} {\it Left:} Moment maps of the central region of
  GDS-14876 showing the projected CO velocity field and velocity
  dispersion where S/N$\geq3$. The white circles marking the extracted
  pseudo slit using apertures with the FWHM of the minimum resolution
  element. The magenta lines illustrate the orientation of the slit in
  the MOSFIRE [OIII] observations, which is misaligned with the CO
  kinematic major axis by $\sim50^{\circ}$. {\it Right:} Observed
  (grey circles) and best-fit models (black line and 1$\sigma$ grey)
  for the rotation velocity and velocity dispersion profiles along the
  kinematic major axis. The dynamical modeling of the kinematic maps
  is consistent with a rotating disk of gas. The dashed lines and grey
  regions indicate the CO effective radius and the 1$\sigma$
  confidence for $r_{\rm e}^{\rm dyn}$.}
\end{figure*}

\section{Kinematics and Dynamical Modeling}

\subsection{[OIII] kinematics}

GDS-14876 was observed using the NIR multi-object spectrograph MOSFIRE
\citep{mosfire1} on Keck-I. A full description of the observations and
data reduction in presented in \citet{barro14b}.  The spectrum yields
a clear ($>5\sigma$) detection of the [OIII] 5007\AA~line (left panel
Figure~\ref{lines}), while the H$\beta$ line is undetected
([OIII]/H$\beta$$\gtrsim0.8$).  The [OIII] profile is relatively
narrow with a $\sigma=55\pm5$~km~s$^{-1}$.

\subsection{CO kinematics}

Figure~\ref{maps} shows the observed rotation velocity and dispersion
fields for GDS-14876 obtained by fitting the CO emission line at every
spaxel with a single Gaussian. The velocity field reveals a continuous
shear and a central dispersion peak which are consistent with the
kinematics of a rotating disk. Assuming that the gas is
gravitationally bound in a disk, we model the observed kinematic
profile to characterize the dynamical properties of the galaxy.

First, we measure the integrated dispersion of the galaxy by fitting a
single Gaussian to the spectrum extracted with a 0\farcs3 diameter
circular aperture. The integrated $\sigma=153\pm7$km~s$^{-1}$ (right
panel of Figure~\ref{lines}) is almost $3\times$ larger than the
dispersion inferred from [OIII]. Such a large difference could
indicate that the ionized and neutral gas trace distinct physical
regions (e.g., if \OIII~traces a relatively unobscured, coronal layer
of the star-forming region).  The misalignment of the MOSFIRE slit
relative to the kinematic major axis (magenta vs. white slit in
leftmost panel of Figure~\ref{maps}) can also lower the
dispersion. However, the linewidth measured on a CO spectrum extracted
along the MOSFIRE slit PA after convolving the ALMA cube to a
FWHM=0\farcs7 resolution still yields a much larger
$\sigma\gtrsim100$km~s$^{-1}$, and suggests that this effect is only
minor.


Next, we analyze the spatially-resolved velocity field using a forward
dynamical modeling procedure which fits the velocity profiles in
observed space. The 1D spectra are extracted along the kinematic major
axis using circular apertures with a diameter equal to the FWHM of the
observations. The dynamical model assumes that the ionized gas is
rotating in a thin disk, and that the disk density profile can be
described by a Freeman model \citep{freeman70}. We account for the
effects of pressure support by lowering the rotation velocity
following \citet[and references therein]{wuyts16}.  The fit includes
the effects of beam smearing (FWHM=0\farcs14$\times$0\farcs12), and
the line-of-sight inclination.  We estimate the latter from the
optical axis-ratio, $q=0.82$, following the method of
\citet{dokkum15}, and we obtain $i=62^{\circ+8}_{-15}$.


\begin{figure}
\centering
\includegraphics[width=7.7cm,bb=37 42 395 407]{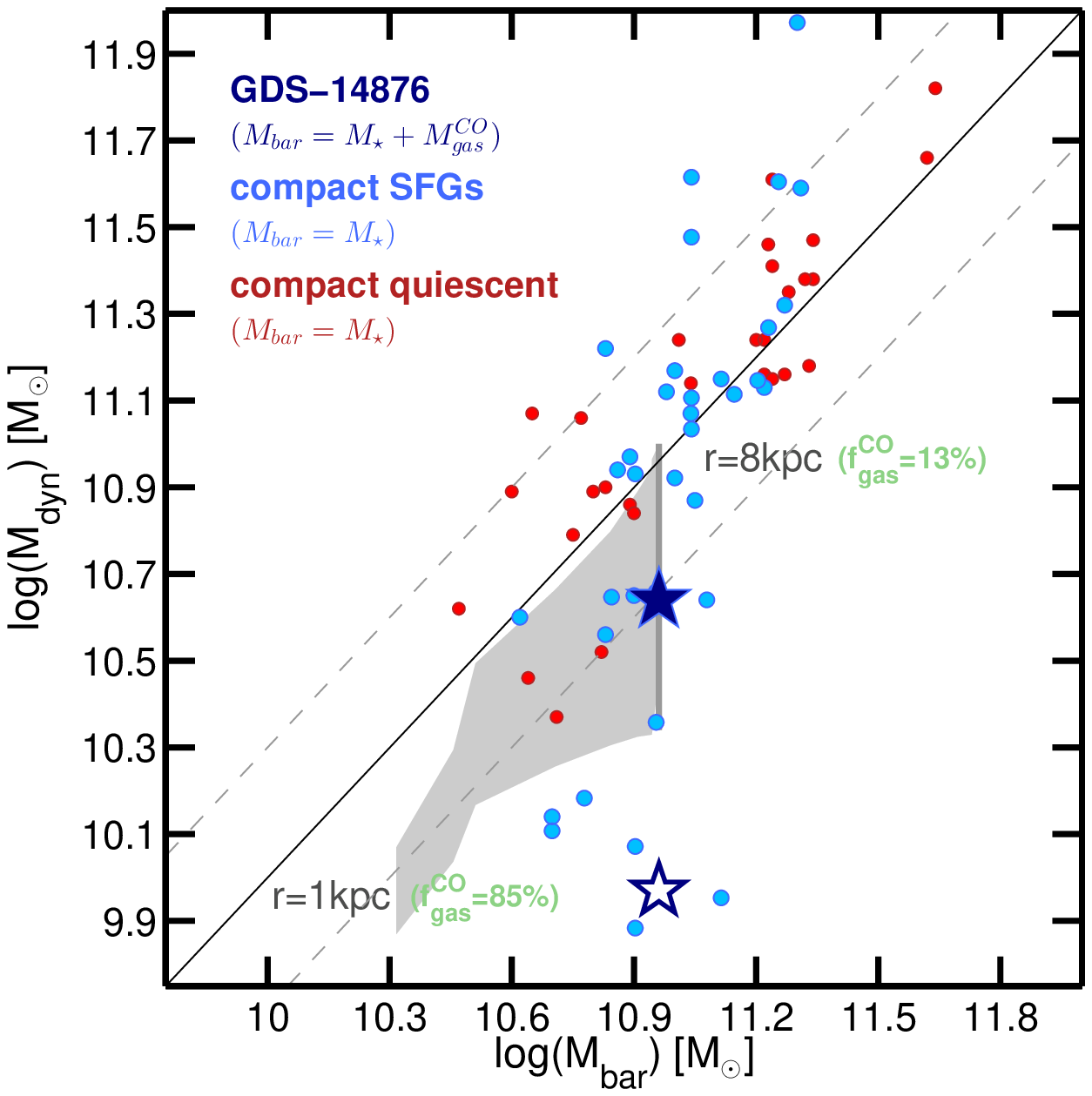}
\caption{\label{dynamics} Comparison between the dynamical and stellar
  masses for GDS-14876 and other samples of compact SFGs (blue) and
  quiescent (red) galaxies from \citet{barro14b}, \citet{dokkum15},
  \citet{vandesande13} and \citet{belli14b}.  The blue stars show
  M$_{\rm dyn}$ for GDS-14876 as computed from the CO (filled) and
  [OIII] (empty) lines. The grey shaded area depicts the confidence
  region of the comparison between cumulative \mdyn~and \mstar with
  increasing radii from $r=1$~kpc to $r=8$~kpc (i.e., total mass). The
  CO-based gas fraction at both ends is indicated. \mdyn~and \mbar~are
  consistent within the confidence range, although the latter lies
  predominantly in the unphysical region, \mbar$>$\mdyn.}
\end{figure}

The free parameters of the dynamical fit are three: the total
dynamical mass and effective radius of the disk, $M_{\rm dyn}$ and
$r^{\rm dyn}_{\rm e}$, and the intrinsic dispersion floor,
$\sigma_{0}$, measured at large radii. Figure~\ref{maps} shows the
observed rotation and dispersion measurements, extracted along a
pseudo slit, compared to the best-fit model.  The best-fit values of
$M_{\rm dyn}$ and $r^{\rm dyn}_{\rm e}$ exhibit a positive correlation
with 1$\sigma$ confidence regions of $r^{\rm dyn}_{\rm e} =
1.7^{+0.5}_{-0.4}$~kpc and \lmdyn=$10.58^{+0.52}_{-0.20}$, where the
largest radius corresponds to the largest mass. Note that the inner
1~kpc is very well constrained by the models.  However, the turnover
radius is only loosely constrained because the CO tracer (magenta),
probes a smaller region than the confidence interval for $r^{\rm
  dyn}_{\rm e}$ (grey region).

The best-fit model yields also an intrinsic
$\sigma_{0}=73\pm8$~km~s$^{-1}$ and a rotation velocity at $r=1$~kpc
of $v_{\rm rot}\sin~i=163\pm5$~km~s$^{-1}$. The dispersion value is
slightly higher but consistent with the typical range observed in
massive SFGs at $z\sim2$ (e.g., \citealt{wisnioski15};
\citealt{price16}; \citealt{tadaki17}). The $v_{\rm
  rot}/\sigma_{0}\sim2.5$, also indicates that the disk is rotation
dominated.

\section{Discussion}

Figure~\ref{dynamics} compares the baryonic (assuming \mgas=0, i.e.,
\mbar=\mstar) and dynamical masses for a sample of massive, compact
SFGs and quiescent galaxies at $z\gtrsim2$ from the literature. Most
galaxies exhibit a relatively good agreement within the usual
$\sim$0.3~dex scatter (dashed line). However, there is a group of
compact SFG outliers with low \mdyn~and \mdyn/\mbar~$\ll1$. These
outliers, found by \citet[][]{dokkum15}, have small integrated
dispersions in their optical emission lines, which leads to small
dynamical masses, $M_{\rm dyn}\propto\sigma^{2}r_{\rm e}$. GDS-14876
exhibits a similar issue, i.e., the dynamical mass computed using
$\sigma$([OIII])=55~km/s is almost 10$\times$ smaller than its
baryonic mass (empty star) even for the smallest fraction of dense,
star-forming gas, $f^{\rm CO}_{\rm gas}\sim13\%$.

The \mdyn~value inferred from the dynamical modeling of the CO
emission line provides a better agreement between total \mdyn~and
\mbar, although the latter is still in the unphysical region (filled
star). As discussed in the previous section, this tension decreases
for larger values of the $r^{\rm dyn}_{\rm e}$ (or the line-of-sight
inclination) which could be better constrained by probing further out
into the rotation curve, possibly by observing lower-order CO
transitions which are sensitive to colder and potentially more
extended neutral gas. Nevertheless, the comparison between total
\mdyn~and \mbar~suggests that the galaxy is strongly baryon dominated,
and the small consistency margin between the two masses favors the
lowest CO-based gas fraction to avoid strongly unphysical results with
\mdyn/\mbar~$\ll1$. A low overall gas fraction also agrees with
similarly low values reported in recent studies of compact SFGs
(\citealt{spilker16}; \citealt{tadaki15}; \citealt{popping17}) and
leaves little room for a colder and more extended gas component that
might be undetected by the high order CO($8-7$) emission.

he spatially-resolved evolution of the $M_{\rm dyn}/M_{\star}(r\leq
R)$ ratio from $R=1$~kpc to $R=8$~kpc exhibits a similar result. The
shaded grey region in Figure~\ref{dynamics} shows that the tension
with \mdyn~increases towards the central region, where the stronger
concentration of star-forming gas relative to stars ($\sim$3$\times$
smaller size) increases the gas fraction to $f^{\rm CO, 1kpc}_{\rm
  gas}\sim85\%$. Note, however, that while dynamical model is better
constrained in the center (narrower grey region) the stellar mass
within 1~kpc is likely more uncertain because the large central
obscuration and limited spatial resolution can both bias the
SED-fitting based stellar mass estimate, particularly for such a
small, compact galaxy.

In summary, the emerging picture from the CO and dust-continuum based
measurements suggests that GDS-14876 is having a compact and likely
short-lived nuclear starburst that could lead to the rapid formation
of a compact quiescent galaxy. A relevant prediction of the kinematic
modeling is that such quiescent descendant could preserve some of the
observed rotational support, as already hinted by recent observations
(e.g., \citealt{newman15a}; \citealt{toft17}).

\section*{Acknowledgments}
GB and MK acknowledge support from HST-AR-12847 and
HST-AR-14552. PGP-G acknowledge support from grants AYA2015-63650-P
and AYA2015-70815-ERC. W.R. is supported by JSPS KAKENHI Grant Number
JP15K17604 and the Thailand Research Fund/Office of the Higher
Education Commission Grant Number MRG6080294. This paper makes use of
the following ALMA data: ADS/JAO.ALMA\#2015.1.00907.S. ALMA is a
partnership of ESO (representing its member states), NSF (USA) and
NINS (Japan), together with NRC (Canada), NSC and ASIAA (Taiwan), and
KASI (Republic of Korea), in cooperation with the Republic of
Chile. The Joint ALMA Observatory is operated by ESO, AUI/NRAO and
NAOJ. The National Radio Astronomy Observatory is a facility of the
National Science Foundation operated under cooperative agreement by
Associated Universities, Inc.


\begin{thebibliography}{54}
\expandafter\ifx\csname natexlab\endcsname\relax\def\natexlab#1{#1}\fi

\bibitem[{{Barro} {et~al.}(2017){Barro}, {Faber}, {Koo}, {Dekel}, {Fang},
  {Trump}, {P{\'e}rez-Gonz{\'a}lez}, {Pacifici}, {Primack}, {Somerville},
  {Yan}, {Guo}, {Liu}, {Ceverino}, {Kocevski}, \& {McGrath}}]{barro17}
{Barro}, G., {Faber}, S.~M., {Koo}, D.~C., {et~al.} 2017, \apj, 840, 47

\bibitem[{{Barro} {et~al.}(2013){Barro}, {Faber}, {P{\'e}rez-Gonz{\'a}lez},
  {Koo}, {Williams}, {Kocevski}, {Trump}, {Mozena}, {McGrath}, {van der Wel},
  {Wuyts}, {Bell}, {Croton}, {Ceverino}, {Dekel}, {Ashby}, {Cheung},
  {Ferguson}, {Fontana}, {Fang}, {Giavalisco}, {Grogin}, {Guo}, {Hathi},
  {Hopkins}, {Huang}, {Koekemoer}, {Kartaltepe}, {Lee}, {Newman}, {Porter},
  {Primack}, {Ryan}, {Rosario}, {Somerville}, {Salvato}, \& {Hsu}}]{barro13}
{Barro}, G., {Faber}, S.~M., {P{\'e}rez-Gonz{\'a}lez}, P.~G., {et~al.} 2013,
  \apj, 765, 104

\bibitem[{{Barro} {et~al.}(2014{\natexlab{a}}){Barro}, {Faber},
  {P{\'e}rez-Gonz{\'a}lez}, {Pacifici}, {Trump}, {Koo}, {Wuyts}, {Guo}, {Bell},
  {Dekel}, {Porter}, {Primack}, {Ferguson}, {Ashby}, {Caputi}, {Ceverino},
  {Croton}, {Fazio}, {Giavalisco}, {Hsu}, {Kocevski}, {Koekemoer},
  {Kurczynski}, {Kollipara}, {Lee}, {McIntosh}, {McGrath}, {Moody},
  {Somerville}, {Papovich}, {Salvato}, {Santini}, {Tal}, {van der Wel},
  {Williams}, {Willner}, \& {Zolotov}}]{barro14a}
---. 2014{\natexlab{a}}, \apj, 791, 52

\bibitem[{{Barro} {et~al.}(2016){Barro}, {Kriek}, {P{\'e}rez-Gonz{\'a}lez},
  {Trump}, {Koo}, {Faber}, {Dekel}, {Primack}, {Guo}, {Kocevski},
  {Mu{\~n}oz-Mateos}, {Rujoparkarn}, \& {Seth}}]{barro16b}
{Barro}, G., {Kriek}, M., {P{\'e}rez-Gonz{\'a}lez}, P.~G., {et~al.} 2016,
  \apjl, 827, L32

\bibitem[{{Barro} {et~al.}(2014{\natexlab{b}}){Barro}, {Trump}, {Koo}, {Dekel},
  {Kassin}, {Kocevski}, {Faber}, {van der Wel}, {Guo},
  {P{\'e}rez-Gonz{\'a}lez}, {Toloba}, {Fang}, {Pacifici}, {Simons}, {Campbell},
  {Ceverino}, {Finkelstein}, {Goodrich}, {Kassis}, {Koekemoer}, {Konidaris},
  {Livermore}, {Lyke}, {Mobasher}, {Nayyeri}, {Peth}, {Primack}, {Rizzi},
  {Somerville}, {Wirth}, \& {Zolotov}}]{barro14b}
{Barro}, G., {Trump}, J.~R., {Koo}, D.~C., {et~al.} 2014{\natexlab{b}}, \apj,
  795, 145

\bibitem[{{Belli} {et~al.}(2014){Belli}, {Newman}, {Ellis}, \&
  {Konidaris}}]{belli14b}
{Belli}, S., {Newman}, A.~B., {Ellis}, R.~S., {et~al.} 2014, \apjl, 788, L29

\bibitem[{{Bradford} {et~al.}(2009){Bradford}, {Aguirre}, {Aikin}, {Bock},
  {Earle}, {Glenn}, {Inami}, {Maloney}, {Matsuhara}, {Naylor}, {Nguyen}, \&
  {Zmuidzinas}}]{bradford09}
{Bradford}, C.~M., {Aguirre}, J.~E., {Aikin}, R., {et~al.} 2009, \apj, 705, 112

\bibitem[{{Bruzual} \& {Charlot}(2003)}]{bc03}
{Bruzual}, G., \& {Charlot}, S. 2003, \mnras, 344, 1000

\bibitem[{{Calzetti} {et~al.}(2000){Calzetti}, {Armus}, {Bohlin}, {Kinney},
  {Koornneef}, \& {Storchi-Bergmann}}]{calzetti}
{Calzetti}, D., {Armus}, L., {Bohlin}, R.~C., {et~al.} 2000, \apj, 533, 682

\bibitem[{{Casey} {et~al.}(2012){Casey}, {Berta}, {B{\'e}thermin}, {Bock},
  {Bridge}, {Budynkiewicz}, {Burgarella}, {Chapin}, {Chapman}, {Clements},
  {Conley}, {Conselice}, {Cooray}, {Farrah}, {Hatziminaoglou}, {Ivison}, {le
  Floc'h}, {Lutz}, {Magdis}, {Magnelli}, {Oliver}, {Page}, {Pozzi},
  {Rigopoulou}, {Riguccini}, {Roseboom}, {Sanders}, {Scott}, {Seymour},
  {Valtchanov}, {Vieira}, {Viero}, \& {Wardlow}}]{casey12}
{Casey}, C.~M., {Berta}, S., {B{\'e}thermin}, M., {et~al.} 2012, \apj, 761, 140

\bibitem[{{Chabrier}(2003)}]{chabrier}
{Chabrier}, G. 2003, \pasp, 115, 763

\bibitem[{{Chary} \& {Elbaz}(2001)}]{ce01}
{Chary}, R., \& {Elbaz}, D. 2001, \apj, 556, 562

\bibitem[{{Daddi} {et~al.}(2015){Daddi}, {Dannerbauer}, {Liu}, {Aravena},
  {Bournaud}, {Walter}, {Riechers}, {Magdis}, {Sargent}, {B{\'e}thermin},
  {Carilli}, {Cibinel}, {Dickinson}, {Elbaz}, {Gao}, {Gobat}, {Hodge}, \&
  {Krips}}]{daddi15}
{Daddi}, E., {Dannerbauer}, H., {Liu}, D., {et~al.} 2015, \aap, 577, A46

\bibitem[{{Daddi} {et~al.}(2005){Daddi}, {Renzini}, {Pirzkal}, {Cimatti},
  {Malhotra}, {Stiavelli}, {Xu}, {Pasquali}, {Rhoads}, {Brusa}, {di Serego
  Alighieri}, {Ferguson}, {Koekemoer}, {Moustakas}, {Panagia}, \&
  {Windhorst}}]{daddi05}
{Daddi}, E., {Renzini}, A., {Pirzkal}, N., {et~al.} 2005, \apj, 626, 680

\bibitem[{{Dale} \& {Helou}(2002)}]{dh02}
{Dale}, D.~A., \& {Helou}, G. 2002, \apj, 576, 159

\bibitem[{{Danielson} {et~al.}(2013){Danielson}, {Swinbank}, {Smail}, {Bayet},
  {van der Werf}, {Cox}, {Edge}, {Henkel}, \& {Ivison}}]{danielson13}
{Danielson}, A.~L.~R., {Swinbank}, A.~M., {Smail}, I., {et~al.} 2013, \mnras,
  436, 2793

\bibitem[{{Draine} \& {Li}(2007)}]{dl07}
{Draine}, B.~T., \& {Li}, A. 2007, \apj, 657, 810

\bibitem[{{Dunne} {et~al.}(2003){Dunne}, {Eales}, {Ivison}, {Morgan}, \&
  {Edmunds}}]{dunne03}
{Dunne}, L., {Eales}, S., {Ivison}, R., {et~al.} 2003, \nat, 424, 285

\bibitem[{{Elbaz} {et~al.}(2011){Elbaz}, {Dickinson}, {Hwang},
  {D{\'{\i}}az-Santos}, {Magdis}, {Magnelli}, {Le Borgne}, {Galliano},
  {Pannella}, {Chanial}, {Armus}, {Charmandaris}, {Daddi}, {Aussel}, {Popesso},
  {Kartaltepe}, {Altieri}, {Valtchanov}, {Coia}, {Dannerbauer}, {Dasyra},
  {Leiton}, {Mazzarella}, {Alexander}, {Buat}, {Burgarella}, {Chary}, {Gilli},
  {Ivison}, {Juneau}, {Le Floc'h}, {Lutz}, {Morrison}, {Mullaney}, {Murphy},
  {Pope}, {Scott}, {Brodwin}, {Calzetti}, {Cesarsky}, {Charlot}, {Dole},
  {Eisenhardt}, {Ferguson}, {F{\"o}rster Schreiber}, {Frayer}, {Giavalisco},
  {Huynh}, {Koekemoer}, {Papovich}, {Reddy}, {Surace}, {Teplitz}, {Yun}, \&
  {Wilson}}]{elbaz11}
{Elbaz}, D., {Dickinson}, M., {Hwang}, H.~S., {et~al.} 2011, \aap, 533, A119

\bibitem[{{Foreman-Mackey} {et~al.}(2013){Foreman-Mackey}, {Hogg}, {Lang}, \&
  {Goodman}}]{mcmc}
{Foreman-Mackey}, D., {Hogg}, D.~W., {Lang}, D., {et~al.} 2013, \pasp, 125, 306

\bibitem[{{Freeman}(1970)}]{freeman70}
{Freeman}, K.~C. 1970, \apj, 160, 811

\bibitem[{{Genzel} {et~al.}(2015){Genzel}, {Tacconi}, {Lutz}, {Saintonge},
  {Berta}, {Magnelli}, {Combes}, {Garc{\'{\i}}a-Burillo}, {Neri}, {Bolatto},
  {Contini}, {Lilly}, {Boissier}, {Boone}, {Bouch{\'e}}, {Bournaud}, {Burkert},
  {Carollo}, {Colina}, {Cooper}, {Cox}, {Feruglio}, {F{\"o}rster Schreiber},
  {Freundlich}, {Gracia-Carpio}, {Juneau}, {Kovac}, {Lippa}, {Naab}, {Salome},
  {Renzini}, {Sternberg}, {Walter}, {Weiner}, {Weiss}, \& {Wuyts}}]{genzel15}
{Genzel}, R., {Tacconi}, L.~J., {Lutz}, D., {et~al.} 2015, \apj, 800, 20

\bibitem[{{Grogin} {et~al.}(2011){Grogin}, {Kocevski}, {Faber}, {Ferguson},
  {Koekemoer}, {Riess}, {Acquaviva}, {Alexander}, {Almaini}, {Ashby}, {Barden},
  {Bell}, {Bournaud}, {Brown}, {Caputi}, {Casertano}, {Cassata}, {Castellano},
  {Challis}, {Chary}, {Cheung}, {Cirasuolo}, {Conselice}, {Roshan Cooray},
  {Croton}, {Daddi}, {Dahlen}, {Dav{\'e}}, {de Mello}, {Dekel}, {Dickinson},
  {Dolch}, {Donley}, {Dunlop}, {Dutton}, {Elbaz}, {Fazio}, {Filippenko},
  {Finkelstein}, {Fontana}, {Gardner}, {Garnavich}, {Gawiser}, {Giavalisco},
  {Grazian}, {Guo}, {Hathi}, {H{\"a}ussler}, {Hopkins}, {Huang}, {Huang},
  {Jha}, {Kartaltepe}, {Kirshner}, {Koo}, {Lai}, {Lee}, {Li}, {Lotz}, {Lucas},
  {Madau}, {McCarthy}, {McGrath}, {McIntosh}, {McLure}, {Mobasher},
  {Moustakas}, {Mozena}, {Nandra}, {Newman}, {Niemi}, {Noeske}, {Papovich},
  {Pentericci}, {Pope}, {Primack}, {Rajan}, {Ravindranath}, {Reddy}, {Renzini},
  {Rix}, {Robaina}, {Rodney}, {Rosario}, {Rosati}, {Salimbeni}, {Scarlata},
  {Siana}, {Simard}, {Smidt}, {Somerville}, {Spinrad}, {Straughn}, {Strolger},
  {Telford}, {Teplitz}, {Trump}, {van der Wel}, {Villforth}, {Wechsler},
  {Weiner}, {Wiklind}, {Wild}, {Wilson}, {Wuyts}, {Yan}, \& {Yun}}]{candelsgro}
{Grogin}, N.~A., {Kocevski}, D.~D., {Faber}, S.~M., {et~al.} 2011, \apjs, 197,
  35

\bibitem[{{Guo} {et~al.}(2013){Guo}, {Ferguson}, {Giavalisco}, {Barro},
  {Willner}, {Ashby}, {Dahlen}, {Donley}, {Faber}, {Fontana}, {Galametz},
  {Grazian}, {Huang}, {Kocevski}, {Koekemoer}, {Koo}, {McGrath}, {Peth},
  {Salvato}, {Wuyts}, {Castellano}, {Cooray}, {Dickinson}, {Dunlop}, {Fazio},
  {Gardner}, {Gawiser}, {Grogin}, {Hathi}, {Hsu}, {Lee}, {Lucas}, {Mobasher},
  {Nandra}, {Newman}, \& {van der Wel}}]{guo13}
{Guo}, Y., {Ferguson}, H.~C., {Giavalisco}, M., {et~al.} 2013, \apjs, 207, 24

\bibitem[{{Kamenetzky} {et~al.}(2016){Kamenetzky}, {Rangwala}, {Glenn},
  {Maloney}, \& {Conley}}]{kamenetzky16}
{Kamenetzky}, J., {Rangwala}, N., {Glenn}, J., {et~al.} 2016, \apj, 829, 93

\bibitem[{{Kellermann} {et~al.}(2008){Kellermann}, {Fomalont}, {Mainieri},
  {Padovani}, {Rosati}, {Shaver}, {Tozzi}, \& {Miller}}]{kellerman08}
{Kellermann}, K.~I., {Fomalont}, E.~B., {Mainieri}, V., {et~al.} 2008, \apjs,
  179, 71

\bibitem[{{Kriek} {et~al.}(2009){Kriek}, {van Dokkum}, {Labb{\'e}}, {Franx},
  {Illingworth}, {Marchesini}, \& {Quadri}}]{fast}
{Kriek}, M., {van Dokkum}, P.~G., {Labb{\'e}}, I., {et~al.} 2009, \apj, 700,
  221

\bibitem[{{McLean} {et~al.}(2010){McLean}, {Steidel}, {Epps}, {Matthews},
  {Adkins}, {Konidaris}, {Weber}, {Aliado}, {Brims}, {Canfield}, {Cromer},
  {Fucik}, {Kulas}, {Mace}, {Magnone}, {Rodriguez}, {Wang}, \&
  {Weiss}}]{mosfire1}
{McLean}, I.~S., {Steidel}, C.~C., {Epps}, H., {et~al.} 2010, in \procspie,
  Vol. 7735, Ground-based and Airborne Instrumentation for Astronomy III,
  77351E--77351E--12

\bibitem[{{McMullin} {et~al.}(2007){McMullin}, {Waters}, {Schiebel}, {Young},
  \& {Golap}}]{mcmullin07}
{McMullin}, J.~P., {Waters}, B., {Schiebel}, D., {et~al.} 2007, in Astronomical
  Society of the Pacific Conference Series, Vol. 376, Astronomical Data
  Analysis Software and Systems XVI, ed. R.~A. {Shaw}, F.~{Hill}, \& D.~J.
  {Bell}, 127

\bibitem[{{Nelson} {et~al.}(2016){Nelson}, {van Dokkum}, {F{\"o}rster
  Schreiber}, {Franx}, {Brammer}, {Momcheva}, {Wuyts}, {Whitaker}, {Skelton},
  {Fumagalli}, {Hayward}, {Kriek}, {Labb{\'e}}, {Leja}, {Rix}, {Tacconi}, {van
  der Wel}, {van den Bosch}, {Oesch}, {Dickey}, \& {Ulf Lange}}]{nelson15}
{Nelson}, E.~J., {van Dokkum}, P.~G., {F{\"o}rster Schreiber}, N.~M., {et~al.}
  2016, \apj, 828, 27

\bibitem[{{Newman} {et~al.}(2015){Newman}, {Belli}, \& {Ellis}}]{newman15a}
{Newman}, A.~B., {Belli}, S., \& {Ellis}, R.~S. 2015, \apjl, 813, L7

\bibitem[{{Peng} {et~al.}(2010){Peng}, {Ho}, {Impey}, \& {Rix}}]{galfit}
{Peng}, C.~Y., {Ho}, L.~C., {Impey}, C.~D., {et~al.} 2010, \aj, 139, 2097

\bibitem[{{P{\'e}rez-Gonz{\'a}lez} {et~al.}(2008){P{\'e}rez-Gonz{\'a}lez},
  {Trujillo}, {Barro}, {Gallego}, {Zamorano}, \& {Conselice}}]{pg08b}
{P{\'e}rez-Gonz{\'a}lez}, P.~G., {Trujillo}, I., {Barro}, G., {et~al.} 2008,
  \apj, 687, 50

\bibitem[{{Popping} {et~al.}(2017){Popping}, {Decarli}, {Man}, {Nelson},
  {B{\'e}thermin}, {De Breuck}, {Mainieri}, {van Dokkum}, {Gullberg}, {van
  Kampen}, {Spaans}, \& {Trager}}]{popping17}
{Popping}, G., {Decarli}, R., {Man}, A.~W.~S., {et~al.} 2017, ArXiv e-prints

\bibitem[{{Price} {et~al.}(2016){Price}, {Kriek}, {Shapley}, {Reddy},
  {Freeman}, {Coil}, {de Groot}, {Shivaei}, {Siana}, {Azadi}, {Barro},
  {Mobasher}, {Sanders}, \& {Zick}}]{price16}
{Price}, S.~H., {Kriek}, M., {Shapley}, A.~E., {et~al.} 2016, \apj, 819, 80

\bibitem[{{Rieke} {et~al.}(2009){Rieke}, {Alonso-Herrero}, {Weiner},
  {P{\'e}rez-Gonz{\'a}lez}, {Blaylock}, {Donley}, \& {Marcillac}}]{rieke09}
{Rieke}, G.~H., {Alonso-Herrero}, A., {Weiner}, B.~J., {et~al.} 2009, \apj,
  692, 556

\bibitem[{{Rujopakarn} {et~al.}(2016){Rujopakarn}, {Dunlop}, {Rieke}, {Ivison},
  {Cibinel}, {Nyland}, {Jagannathan}, {Silverman}, {Alexander}, {Biggs},
  {Bhatnagar}, {Ballantyne}, {Dickinson}, {Elbaz}, {Geach}, {Hayward},
  {Kirkpatrick}, {McLure}, {Micha{\l}owski}, {Miller}, {Narayanan}, {Owen},
  {Pannella}, {Papovich}, {Pope}, {Rau}, {Robertson}, {Scott}, {Swinbank}, {van
  der Werf}, {van Kampen}, {Weiner}, \& {Windhorst}}]{rujopakarn16}
{Rujopakarn}, W., {Dunlop}, J.~S., {Rieke}, G.~H., {et~al.} 2016, \apj, 833, 12

\bibitem[{{Sandstrom} {et~al.}(2013){Sandstrom}, {Leroy}, {Walter}, {Bolatto},
  {Croxall}, {Draine}, {Wilson}, {Wolfire}, {Calzetti}, {Kennicutt}, {Aniano},
  {Donovan Meyer}, {Usero}, {Bigiel}, {Brinks}, {de Blok}, {Crocker}, {Dale},
  {Engelbracht}, {Galametz}, {Groves}, {Hunt}, {Koda}, {Kreckel}, {Linz},
  {Meidt}, {Pellegrini}, {Rix}, {Roussel}, {Schinnerer}, {Schruba}, {Schuster},
  {Skibba}, {van der Laan}, {Appleton}, {Armus}, {Brandl}, {Gordon}, {Hinz},
  {Krause}, {Montiel}, {Sauvage}, {Schmiedeke}, {Smith}, \&
  {Vigroux}}]{sandstrom13}
{Sandstrom}, K.~M., {Leroy}, A.~K., {Walter}, F., {et~al.} 2013, \apj, 777, 5

\bibitem[{{Scoville} {et~al.}(2016){Scoville}, {Sheth}, {Aussel}, {Vanden
  Bout}, {Capak}, {Bongiorno}, {Casey}, {Murchikova}, {Koda},
  {{\'A}lvarez-M{\'a}rquez}, {Lee}, {Laigle}, {McCracken}, {Ilbert}, {Pope},
  {Sanders}, {Chu}, {Toft}, {Ivison}, \& {Manohar}}]{scoville16}
{Scoville}, N., {Sheth}, K., {Aussel}, H., {et~al.} 2016, \apj, 820, 83

\bibitem[{{Spilker} {et~al.}(2016){Spilker}, {Bezanson}, {Marrone}, {Weiner},
  {Whitaker}, \& {Williams}}]{spilker16}
{Spilker}, J.~S., {Bezanson}, R., {Marrone}, D.~P., {et~al.} 2016, \apj, 832,
  19

\bibitem[{{Tacconi} {et~al.}(2008){Tacconi}, {Genzel}, {Smail}, {Neri},
  {Chapman}, {Ivison}, {Blain}, {Cox}, {Omont}, {Bertoldi}, {Greve},
  {F{\"o}rster Schreiber}, {Genel}, {Lutz}, {Swinbank}, {Shapley}, {Erb},
  {Cimatti}, {Daddi}, \& {Baker}}]{tacconi08}
{Tacconi}, L.~J., {Genzel}, R., {Smail}, I., {et~al.} 2008, \apj, 680, 246

\bibitem[{{Tadaki} {et~al.}(2017){Tadaki}, {Kodama}, {Nelson}, {Belli},
  {F{\"o}rster Schreiber}, {Genzel}, {Hayashi}, {Herrera-Camus}, {Koyama},
  {Lang}, {Lutz}, {Shimakawa}, {Tacconi}, {{\"U}bler}, {Wisnioski}, {Wuyts},
  {Hatsukade}, {Lippa}, {Nakanishi}, {Ikarashi}, {Kohno}, {Suzuki}, {Tamura},
  \& {Tanaka}}]{tadaki17}
{Tadaki}, K.-i., {Kodama}, T., {Nelson}, E.~J., {et~al.} 2017, \apjl, 841, L25

\bibitem[{{Tadaki} {et~al.}(2015){Tadaki}, {Kohno}, {Kodama}, {Ikarashi},
  {Aretxaga}, {Berta}, {Caputi}, {Dunlop}, {Hatsukade}, {Hayashi}, {Hughes},
  {Ivison}, {Izumi}, {Koyama}, {Lutz}, {Makiya}, {Matsuda}, {Nakanishi},
  {Rujopakarn}, {Tamura}, {Umehata}, {Wang}, {Wilson}, {Wuyts}, {Yamaguchi}, \&
  {Yun}}]{tadaki15}
{Tadaki}, K.-i., {Kohno}, K., {Kodama}, T., {et~al.} 2015, \apjl, 811, L3

\bibitem[{{Toft} {et~al.}(2017){Toft}, {Zabl}, {Richard}, {Gallazzi},
  {Zibetti}, {Prescott}, {Grillo}, {Man}, {Lee}, {G{\'o}mez-Guijarro},
  {Stockmann}, {Magdis}, \& {Steinhardt}}]{toft17}
{Toft}, S., {Zabl}, J., {Richard}, J., {et~al.} 2017, \nat, 546, 510

\bibitem[{{Trujillo} {et~al.}(2007){Trujillo}, {Conselice}, {Bundy}, {Cooper},
  {Eisenhardt}, \& {Ellis}}]{trujillo07}
{Trujillo}, I., {Conselice}, C.~J., {Bundy}, K., {et~al.} 2007, \mnras, 382,
  109

\bibitem[{{van de Sande} {et~al.}(2013){van de Sande}, {Kriek}, {Franx}, {van
  Dokkum}, {Bezanson}, {Bouwens}, {Quadri}, {Rix}, \& {Skelton}}]{vandesande13}
{van de Sande}, J., {Kriek}, M., {Franx}, M., {et~al.} 2013, \apj, 771, 85

\bibitem[{{van Dokkum} {et~al.}(2015){van Dokkum}, {Nelson}, {Franx}, {Oesch},
  {Momcheva}, {Brammer}, {F{\"o}rster Schreiber}, {Skelton}, {Whitaker}, {van
  der Wel}, {Bezanson}, {Fumagalli}, {Illingworth}, {Kriek}, {Leja}, \&
  {Wuyts}}]{dokkum15}
{van Dokkum}, P.~G., {Nelson}, E.~J., {Franx}, M., {et~al.} 2015, \apj, 813, 23

\bibitem[{{Wellons} {et~al.}(2015){Wellons}, {Torrey}, {Ma}, {Rodriguez-Gomez},
  {Vogelsberger}, {Kriek}, {van Dokkum}, {Nelson}, {Genel}, {Pillepich},
  {Springel}, {Sijacki}, {Snyder}, {Nelson}, {Sales}, \&
  {Hernquist}}]{wellons15}
{Wellons}, S., {Torrey}, P., {Ma}, C.-P., {et~al.} 2015, \mnras, 449, 361

\bibitem[{{Whitaker} {et~al.}(2014){Whitaker}, {Franx}, {Leja}, {van Dokkum},
  {Henry}, {Skelton}, {Fumagalli}, {Momcheva}, {Brammer}, {Labb{\'e}},
  {Nelson}, \& {Rigby}}]{whitaker14}
{Whitaker}, K.~E., {Franx}, M., {Leja}, J., {et~al.} 2014, \apj, 795, 104

\bibitem[{{Wisnioski} {et~al.}(2015){Wisnioski}, {F{\"o}rster Schreiber},
  {Wuyts}, {Wuyts}, {Bandara}, {Wilman}, {Genzel}, {Bender}, {Davies},
  {Fossati}, {Lang}, {Mendel}, {Beifiori}, {Brammer}, {Chan}, {Fabricius},
  {Fudamoto}, {Kulkarni}, {Kurk}, {Lutz}, {Nelson}, {Momcheva}, {Rosario},
  {Saglia}, {Seitz}, {Tacconi}, \& {van Dokkum}}]{wisnioski15}
{Wisnioski}, E., {F{\"o}rster Schreiber}, N.~M., {Wuyts}, S., {et~al.} 2015,
  \apj, 799, 209

\bibitem[{{Wuyts} {et~al.}(2012){Wuyts}, {F{\"o}rster Schreiber}, {Genzel},
  {Guo}, {Barro}, {Bell}, {Dekel}, {Faber}, {Ferguson}, {Giavalisco}, {Grogin},
  {Hathi}, {Huang}, {Kocevski}, {Koekemoer}, {Koo}, {Lotz}, {Lutz}, {McGrath},
  {Newman}, {Rosario}, {Saintonge}, {Tacconi}, {Weiner}, \& {van der
  Wel}}]{wuyts12}
{Wuyts}, S., {F{\"o}rster Schreiber}, N.~M., {Genzel}, R., {et~al.} 2012, \apj,
  753, 114

\bibitem[{{Wuyts} {et~al.}(2011){Wuyts}, {F{\"o}rster Schreiber}, {van der
  Wel}, {Magnelli}, {Guo}, {Genzel}, {Lutz}, {Aussel}, {Barro}, {Berta},
  {Cava}, {Graci{\'a}-Carpio}, {Hathi}, {Huang}, {Kocevski}, {Koekemoer},
  {Lee}, {Le Floc'h}, {McGrath}, {Nordon}, {Popesso}, {Pozzi}, {Riguccini},
  {Rodighiero}, {Saintonge}, \& {Tacconi}}]{wuyts11b}
{Wuyts}, S., {F{\"o}rster Schreiber}, N.~M., {van der Wel}, A., {et~al.} 2011,
  \apj, 742, 96

\bibitem[{{Wuyts} {et~al.}(2016){Wuyts}, {F{\"o}rster Schreiber}, {Wisnioski},
  {Genzel}, {Burkert}, {Bandara}, {Beifiori}, {Belli}, {Bender}, {Brammer},
  {Chan}, {Davies}, {Fossati}, {Galametz}, {Kulkarni}, {Lang}, {Lutz},
  {Mendel}, {Momcheva}, {Naab}, {Nelson}, {Saglia}, {Seitz}, {Tacconi},
  {Tadaki}, {{\"U}bler}, {van Dokkum}, {Wilman}, \& {Wuyts}}]{wuyts16}
{Wuyts}, S., {F{\"o}rster Schreiber}, N.~M., {Wisnioski}, E., {et~al.} 2016,
  \apj, 831, 149

\bibitem[{{Zolotov} {et~al.}(2015){Zolotov}, {Dekel}, {Mandelker}, {Tweed},
  {Inoue}, {DeGraf}, {Ceverino}, {Primack}, {Barro}, \& {Faber}}]{zolotov15}
{Zolotov}, A., {Dekel}, A., {Mandelker}, N., {et~al.} 2015, \mnras, 450, 2327

\end{thebibliography}

\end{document}